\documentclass{aa}
\usepackage{graphicx}
\usepackage{times,epsfig,amssymb,amsmath}

\def\ltsima{$\; \buildrel < \over \sim \;$}
\def\simlt{\lower.5ex\hbox{\ltsima}}
\def\gtsima{$\; \buildrel > \over \sim \;$}
\def\simgt{\lower.5ex\hbox{\gtsima}}

\begin{document}

\vskip 2.truecm
   \title{On the Iron content in rich nearby Clusters of Galaxies} 

   \author{S. De Grandi \inst{1}, S. Ettori \inst{2}, M. Longhetti
   \inst{1} \and S. Molendi \inst{3} }

   \offprints{S. De Grandi}

   \institute{INAF - Osservatorio Astronomico di Brera,
              via E. Bianchi 46, I-23807 Merate (LC)\\
              \email{degrandi@mi.astro.it; marcella@mi.astro.it}
         \and
             ESO - Headquarter,
             Karl-Schwarzschild-Str. 2, D-85748 Garching bei M\"unchen\\
             \email{settori@eso.org}
         \and
             Istituto di Astrofisica Spaziale e Fisica Cosmica (CNR)
             via Bassini 15, I-20133 Milano\\
             \email{silvano@mi.iasf.cnr.it}
             }

   \date{Accepted for publication in A\&A }

   \abstract{

In this paper we study the iron content of a sample of 22 nearby hot
clusters observed with {\it BeppoSAX}.  We find that the global iron
mass of clusters is tightly related to the cluster luminosity and that
the relatively loose correlation between the iron mass and the cluster
temperature follows from the combination of the iron mass vs.
luminosity and luminosity vs.  temperature correlations.  The iron
mass is found to scale linearly with the intracluster gas mass,
implying that the global iron abundance in clusters is roughly
constant.  This result suggests that enrichment mechanisms operate at
a similar rate in all clusters.  By employing population synthesis and
chemical enrichment models, we show that the iron mass associated to
the abundance excess which is always found in the centre of cool core
clusters can be entirely produced by the brightest cluster galaxy
(BCG), which is always found at the centre of cool core clusters.  The
iron mass associated to the excess, the optical magnitude of the BCG
and the temperature of the cluster are found to correlate with one
another suggesting a link between the properties of the BCG and the
hosting cluster.  These observational facts lends strength to current
formation theories which envisage a strong connection between the
formation of the giant BCG and its hosting cluster.

    }

   \titlerunning {On the     Iron content in   Clusters  of  Galaxies}
   \authorrunning {De Grandi, Ettori, Longhetti, Molendi}

   \maketitle

\section{Introduction}
The X-ray  thermal     continuum observed in clusters    of   galaxies
originates   from the hot intra-cluster   medium  (ICM) permeating the
cluster potential well.  At  the  high temperatures measured in   rich
clusters, kT$\simgt 3$ keV, the ICM is highly ionised and its spectrum
presents a number of emission lines, among which the most prominent is
the  Fe   K-shell line at  $\sim 7$   keV.  From  the   measure of the
equivalent  width of these  spectral lines it  is possible to estimate
the abundance of a  given element.  A  well  known result is that  the
global iron abundance in clusters is about  a third of the solar value
and that   this  value remains  constant  within  redshift  $\simlt 1$
(Mushotzky  \& Loewenstein 1997, Allen \&  Fabian  1998, Della Ceca et
al. 2000, Ettori, Allen \& Fabian 2001, Tozzi et al. 2003), indicating
that the  enrichment process has mostly taken  place at early times in
the cluster formation.

The ICM iron  mass is a key observable  to constrain the integral past
star  formation  history  in clusters, and  its    relation with other
observables such as the cluster optical light,  total cluster mass and
stellar  mass allows  us to  study  the  enrichment processes and  the
related efficiencies.   The iron mass  of the ICM correlates  with the
optical  light of massive early-type   galaxies in clusters (Arnaud et
al. 1992), suggesting that the stellar population of these systems are
the main responsible of   the ICM enrichment.  Further evidences  that
clusters evolved experiencing  very  similar star  formation histories
comes from   the observed constancy  of  the  iron-mass-to-light ratio
(Ciotti et al.  1991, Renzini et al.  1993,  Renzini 1997), and of the
iron-mass-to-total-mass ratio    (Lin, Mohr \&   Stanford 2003)  among
clusters.  These evidences, and   the similar global   abundance among
clusters, support the  possibility   that  rich clusters evolved    as
"closed boxes" after their formation, with little baryon exchange with
the surroundings  (Renzini  1997).  While   the  origin of  the metals
observed   in  the   ICM is clearly     related  to  stellar evolution
(supernovae events), the transfer mechanism of these metals from stars
in   galaxies to  the  ICM  is less   clear.   Possible ICM enrichment
mechanisms  that have  been proposed  for clusters   are: ram pressure
stripping of  metal enriched gas  from cluster galaxies (e.g.  Gunn \&
Gott 1972, Toniazzo \& Schindler 2001), and  stellar winds AGN- or SN-
induced   in  early-type   galaxies   during  the  formation   of  the
proto-cluster (e.g.  Gnedin 1998, Kauffmann  \& Charlot 1998, Dupke \&
White 2000,  Renzini et al.  1993,  Lin et al.  2003).  The efficiency
of the metal extraction by ram pressure stripping should increase with
the local density of galaxies,  the lack of any   trend of either  the
iron  abundance and iron-mass-to-light  ratio with cluster temperature
or velocity  dispersion argues against  this being the main mechanisms
of metal transportation at work in clusters (Renzini 2003).

Spatially   resolved analysis of   metal  abundances in  clusters have
become  possible  only  recently,   first  with  {\it ASCA}  and  {\it
BeppoSAX} and, more recently, with {\it Chandra} and {\it XMM-Newton}.
These measurements have shown  that abundance gradients are common  in
clusters and groups  of galaxies ({\it ASCA}: Fukazawa  et  al.  2000,
Makishima  et  al.  2001,  Finoguenov,   Arnaud  \&  David 2001;  {\it
BeppoSAX}:   Irwin \& Bregman 2001,  De  Grandi \&  Molendi 2001; {\it
Chandra}: Ettori et al.  2002, Hicks  et al.  2002, Schmidt, Fabian \&
Sanders   2002, Blanton et  al.  2003;  {\it   XMM-Newton}: Molendi \&
Gastaldello 2001, Gastaldello  \&   Molendi 2002, Finoguenov  et   al.
2002, Pratt \&   Arnaud 2002,  Matsushita,  Finoguenov  \&  Boehringer
2003).  Relaxed clusters with evidences    of cool cores have   global
abundance about twice larger than the abundance of non relaxed systems
(Allen \& Fabian 1998).  De Grandi \&  Molendi (2001, DM01 hereafter),
found that this is due to  the almost ubiquitous presence of abundance
peaks  in the cores   of  relaxed clusters  in correspondence   of the
brightest cluster galaxy   (BCG), while clusters showing evidences  of
dynamical activity on   large scales  have  flat  abundance  profiles.
Comparison between the  total  cluster  light and  abundance  profiles
suggests that the abundance peaks are probably due to the accumulation
of  metal ejection from  the BCG  in the ICM  (Fukazawa et  al.  2000,
DM01).

The  plan of  the paper is  as follows:  in   Sect. 2 we  describe the
sample,  give a brief  review on   the data  analysis and  discuss the
deprojected  abundance profiles. In Sect.  3  we compute  the ICM iron
mass for our  clusters and study how this  mass scales with  the other
important physical observables. We estimate the metal content of local
clusters and  compare it to the cosmological  metal budget. In Sect. 4
we investigate the origin of  the iron mass  excess associated to  the
abundance    peak  observed in cool   core   clusters and its possible
connection with the brightest cluster  galaxy located at the centre of
these clusters.  Finally we summarise our findings in Sect. 5.

In  the rest of the  paper we employ  the  solar abundance ratios from
Grevesse  \& Sauval (1998) with $Z_\odot={\rm Fe/H}=3.16\times10^{-5}$
by  number,   as  suggested by Brighenti    \& Mathews   (1999).  This
"standard  solar  composition" derived  by  Grevesse  \& Sauval (1998)
follows from recent  photospheric models of  the Sun and shows perfect
agreement with the meteoritic composition. We recall here that the new
abundances can be  easily converted into  abundances relative to other
sets of  solar  abundances with a  simple  scaling. The multiplicative
factor necessary  to convert abundances  relative to solar values from
Grevesse \& Sauval  (1998)  to those   relative to  solar  values from
Anders \& Grevesse (1989) is 0.675.

The cosmological framework is an Einstein-de Sitter model with H$_0$ =
50 Mpc  km s$^{-1}$ to  allow comparison with  previous works.  Quoted
confidence  intervals are  $68\%$ for  1  interesting parameter  (i.e.
$\Delta \chi^2 =1$), unless otherwise stated.

\section {The Sample and deprojected Abundance Profiles}
We consider a sample of 22 nearby ($z\simlt 0.1$) clusters of galaxies
observed with the  two Medium Energy Concentrator Spectrometers (MECS)
on  board   {\it BeppoSAX}.   These  clusters cover  the  range in gas
temperature between 3 and 10 keV, and have bolometric X-ray luminosity
between $2\times   10^{44}$ erg s$^{-1}$  and $6   \times 10^{45}$ erg
s$^{-1}$.  We have divided the sample  into a subsample of 10 clusters
without evidences of cool cores (NCC clusters hereafter, these objects
were called non-CF clusters  in    our previous papers)  and   another
subsample of 12 clusters with cool cores  (CC clusters hereafter, they
were called CF clusters in our previous  papers).  We have defined NCC
cluster  an object  with a mass  deposition rate  consistent with zero
according to the     spatial  analysis of   {\it   ROSAT} observations
presented in Peres  et al.  (1998).    Contrary to NCC  systems, which
show  signatures of  recent merging  events,  CC  clusters are  mostly
relaxed systems (at  least on the scale  of 100 kpc) with  evidence of
strong emission peaks (e.g.  Mohr,  Mathiesen \& Evrard 1999) and cool
gas  in  their centres (e.g.  Peres   et al.  1998,  Allen, Schmidt \&
Fabian 2001).

Details on the  data analysis and  deprojection technique are given in
the previous papers of this series, DM01,  De Grandi \& Molendi (2002)
and Ettori, De Grandi  \& Molendi (2002) (EDM02  in  the rest of  this
paper).  The observation  log for the current  cluster sample is given
in  Table 1 of  EDM02.  The observed   metal abundance and temperature
profiles have   been discussed in details  in  DM01 and De   Grandi \&
Molendi   (2002), respectively.     We  remark  here that  the   metal
abundances    determined  for these   clusters  are  essentially  iron
abundances  derived from the K-shell iron  line at $\sim 6.9$ keV, and
that the other elements in the ICM are scaled in solar proportion.  In
EDM02 we  have  converted the  physical  quantities  derived from  the
fitting of the MECS spectra  extracted from cluster regions  projected
on   the sky, into their deprojected    values under the assumption of
spherical geometry of the  X-ray emitting plasma.   For the first time
these {\it BeppoSAX}   observations allowed us  to obtain  deprojected
physical quantities of clusters from one dataset only.

%----------------------------------------------------------------------
\begin{figure*}
\hbox{  
\hskip -0.5cm
\epsfig{figure=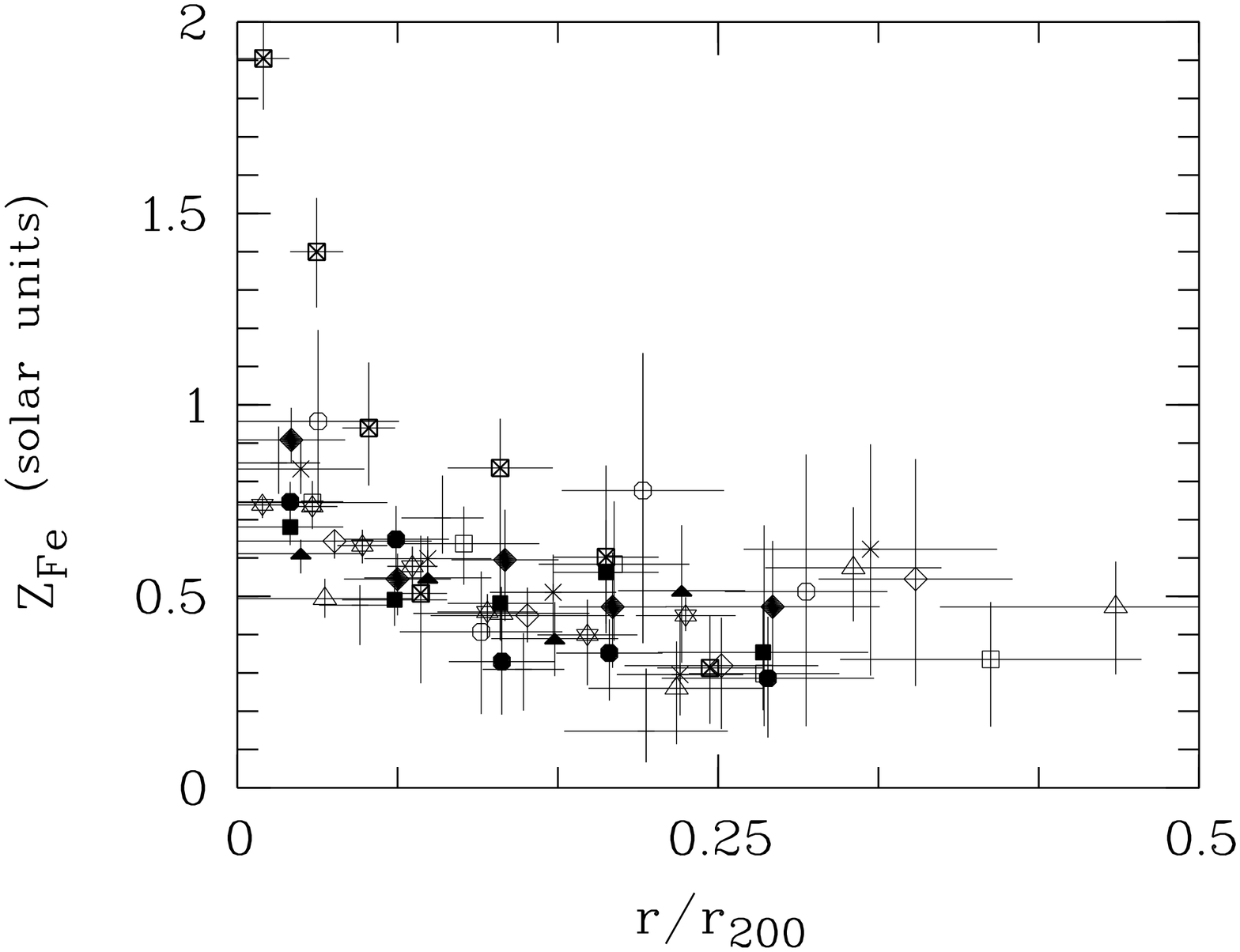,width=0.37\textwidth,angle=-90}  
\hskip -0.8cm
\epsfig{figure=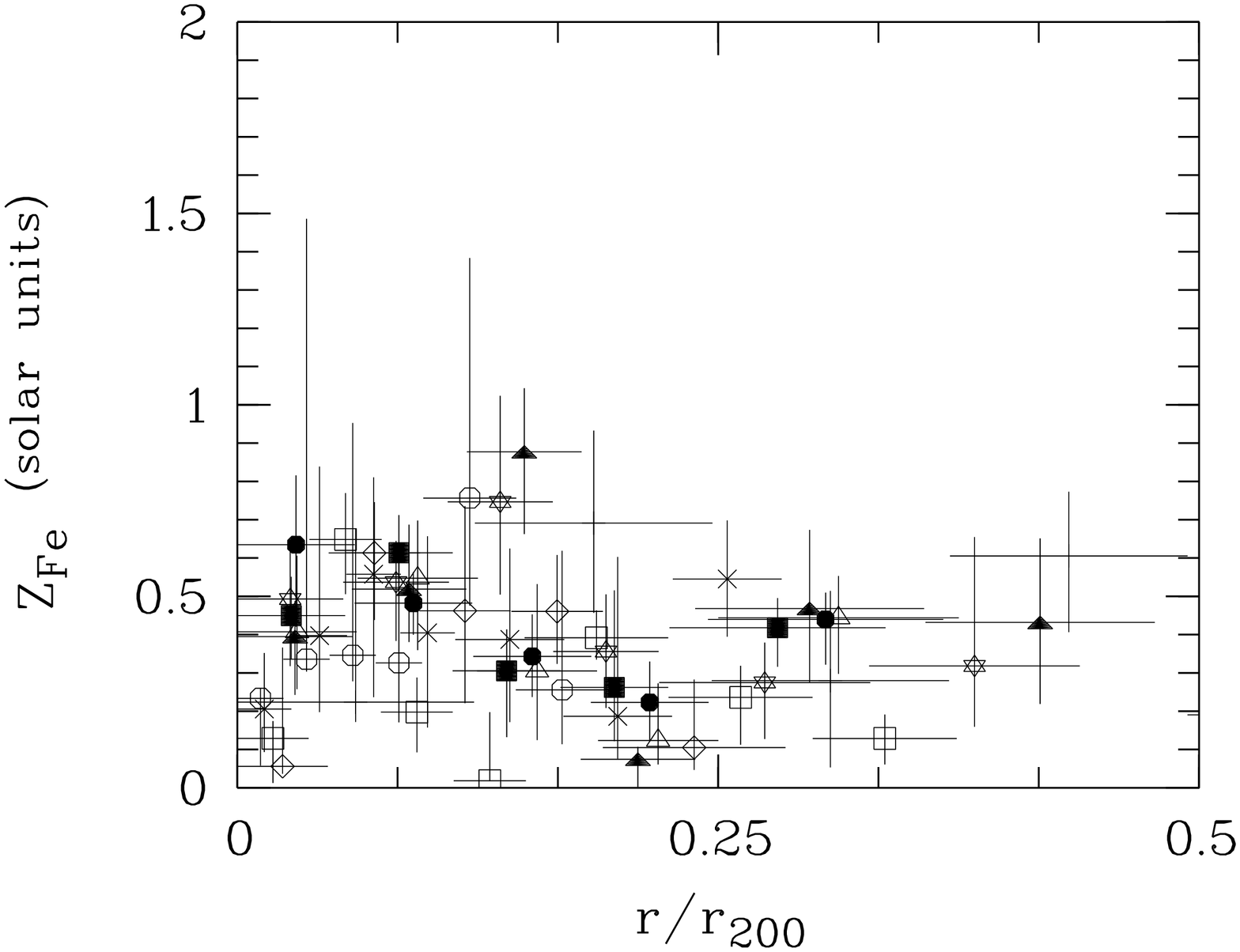,width=0.37\textwidth,angle=-90}  
} 
\caption{ Deprojected iron abundance profiles for CC ({\it left})
and NCC ({\it  right}) clusters  plotted as  a function  of the radius
normalised to $r_{200}$.  Clusters are  related to symbols as follows:
({\it left panel}) A85 ({\it filled circles}), A426 ({\it asterisks}),
A496 ({\it filled lozenges}),  A1795  ({\it filled triangles}),  A2029
({\it open squares}), A2142 ({\it open triangles}), A2199 ({\it filled
squares}), A3526 ({\it crossed squares}),  A3562 ({\it open circles}),
A3571 ({\it crosses $+$}), 2A 0335$+$096 ({\it crosses $\times$}), and
PKS 0745$-$191  ({\it open lozenges});  ({\it right panel}) A119 ({\it
open    lozenges}),  A754 ({\it filled   squares}),   A1367 ({\it open
circles}), A1656 ({\it  open squares}), A2256 ({\it  filled circles}),
A2319 ({\it  filled triangles}),  A3266 ({\it open  triangles}), A3376
({\it crosses $+$}),   A3627 ({\it crosses  $\times$}), and Triangulum
Austr. ({\it asterisks}).  }
\label{Fig1} 
\end{figure*}  
%----------------------------------------------------------------------

The projected (observed) metal  abundance  profiles presented in  DM01
provided evidences that CC clusters have strong abundance enhancements
in their cores with respect to NCC systems, which on the contrary tend
to have  constant   metallicities.  In   Fig.~\ref{Fig1} we show   the
deprojected iron  abundance   profiles for both  cluster   types, as a
function of the radius  normalised  to $r_{200}$ (values of  $r_{200}$
are given in EDM02).  The comparison  between the deprojected profiles
is in agreement with the comparison of  projected profiles reported in
DM01: NCC systems have  almost  flat metallicity profiles whereas   CC
ones have  metallicity enhancements in  the core and  flat profiles in
the outermost regions.  The best-fit value of  the metallicity using a
constant  model over the  whole   radial range for    both CC and  NCC
clusters  is $0.62\pm  0.01$  and  $0.32\pm 0.01$,  respectively.  The
constant model    however    provides  a poor   fit    in  both  cases
($\chi^2$/dof=302/60 for CC and 112/56 for NCC clusters).

In the case of NCC clusters the fact that the  constant model does not
reproduce the data adequately  is due to  the  large scatter in  their
deprojected profiles. We  recall  that  the deprojected profiles   are
obtained by applying a  deprojection technique which assumes  that the
sources are spherically symmetric, in the case of NCC clusters this is
a rather strong simplification and may result in the observed scatter.

In the case of CC clusters the constant model is clearly a poor fit to
the profiles.  The best-fit to  the data with a $\beta$-model, $Z_{\rm
Fe}=Z_{\rm  Fe   0}(1+(x/x_c))^{-\alpha}$  where $x=r/r_{200}$,  gives
$Z_{\rm  Fe  0}=0.80\pm  0.03$,    $x_{\rm    c}=0.04\pm  0.01$    and
$\alpha=0.18\pm 0.03$ ($\chi^2$/dof=140/58), an  F-test shows that the
improvement with  respect to the constant  model is highly significant
(P$_F>99.99\%$).  Note that  the   metallicity profile  of   Centaurus
cluster   (A3526) is  peculiar  in the   sense  that the   deprojected
abundance at its centre is the only one,  among our CC clusters, to be
larger  than  solar.   By excluding  Centaurus both   the constant and
$\beta$  models fit on the   remaining  CC  clusters give   parameters
indistinguishable from those obtained considering all CC clusters. The
only important  difference  is that, when  Centaurus is  excluded, the
$\beta$-model becomes acceptable ($\chi^2$/dof=69/51).

%----------------------------------------------------------------------
\begin{figure}
\begin{center}
\epsfig{figure=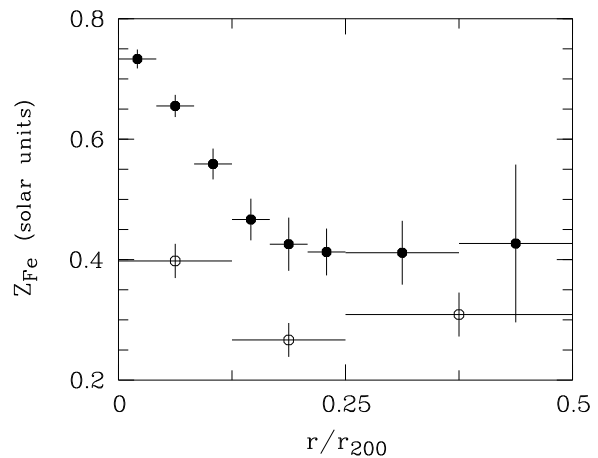,width=0.35\textwidth,angle=-90}  
\caption{ Binned weighted average of the deprojected metallicity profiles 
for CC ({\it filled circles}) and NCC ({\it open circles}) clusters.}
\label{Fig2} 
\end{center} 
\end{figure}
%----------------------------------------------------------------------

In Fig.~\ref{Fig2} we  plot, for both CC and  NCC clusters, the binned
error-weighted   average  of   the   metallicity   profiles shown   in
Fig.~\ref{Fig1}.  From this figure we note  that the mean abundance at
large  radii is  different    between  CC and   NCC    clusters.   The
best-fitting constant  to    the mean profile   of  CC   clusters  for
$r>0.2~r_{200}$  gives    an average   abundance   of $0.40\pm   0.03$
($\chi^2$/dof=12.4/17),  which differs by $\sim   2.2 \sigma$ from the
average  abundance over  the   whole  radial range   of  NCC  clusters
($0.32\pm 0.01$).    Although the   statistical significance  of  this
difference is small  we  suggest  that this   could indicate that   CC
clusters  are older systems  than  the  NCC ones.   In  this case  the
early-type  galaxy population of CC  clusters had more time to produce
metals and eject them into  the ICM with respect to  NCC clusters.  An
independent indication that relaxed clusters  are probably older  than
disturbed systems, comes from the work of Katayama et al.  (2003), who
found that the offset of the brightest cluster galaxy from the peak of
cluster  X-ray emission is   larger for younger  (i.e.   less relaxed)
clusters.  New {\it XMM-Newton}  measurements of cluster abundances in
the outermost regions  of the clusters  are  required to make  further
progress on this important issue.

%----------------------------------------------------------------------  
\begin{table*}
\begin{center}
\caption{Sample of 22 galaxy clusters considered in this study. 
The `CC' column indicates if a cluster is or is not a cool core object
according to the mass deposition rate quoted  in Peres et al.  (1998).
The ICM iron mass computed at  different overdensities are reported in
the third, fourth and fifth columns. The sixth column reports the iron
mass excess measured for CC clusters only.  }
\label{tab1}
\begin{tabular}{l@{\hspace{.8em}} c@{\hspace{.6em}} c@{\hspace{.8em}}
c@{\hspace{.8em}} c@{\hspace{.8em}} c@{\hspace{.8em}}} 
 Cluster & CC & ${\rm M_{Fe}}$ & ${\rm M_{Fe}}$ & 
 ${\rm M_{Fe}^{(\dagger)}}$ & ${\rm M_{Fe}^{exc}}$  \\
 &  & ${\rm 10^{10} M_{\odot}}$ & ${\rm 10^{10} M_{\odot}}$ & 
 ${\rm 10^{10} M_{\odot}}$ & ${\rm 10^9 M_{\odot}}$ \\
 & & $\Delta=2500$ & $\Delta=1000$ & $\Delta=500$
 \\ \\ 
A85             & y & 3.90(1.06) & 7.83(1.90) & 12.1(2.80) & 4.10(1.92) \\ 
A426            & y & 2.90(0.31) & 4.94(0.47) & 7.10(0.63) & 3.57(1.43) \\ 
A496            & y & 1.98(0.41) & 3.53(0.65) & 5.09(0.89) & 1.58(0.54) \\ 
A1795           & y & 3.73(0.59) & 8.00(1.17) & 12.7(1.80) & 3.28(1.69) \\ 
A2029           & y & 6.00(1.96) & 12.0(3.34) & 18.7(4.77) & 11.6(5.63) \\ 
A2142           & y & 5.59(1.47) & 13.3(3.04) & 21.9(4.21) & 1.09(0.96) \\ 
A2199           & y & 1.63(0.30) & 2.69(0.46) & 3.76(0.62) & 1.50(0.95) \\ 
A3526           & y & 0.54(0.12) & 0.88(0.17) & 1.23(0.22) & 0.64(0.21) \\ 
A3562           & y & 1.77(0.65) & 3.72(1.10) & 5.67(1.54) & 0.61(0.34) \\ 
A3571           & y & 3.07(0.66) & 5.34(1.15) & 7.77(1.68) & 3.81(1.96) \\ 
2A 0335$+$096   & y & 1.61(0.36) & 3.01(0.64) & 4.41(0.92) & 1.77(0.54) \\ 
PKS 0745$-$191  & y & 5.49(1.23) & 12.3(2.69) & 20.1(4.22) & 4.89(1.43) \\ 
A119            & n & 1.38(0.52) & 2.59(0.86) & 3.76(1.19) & ---        \\ 
A754            & n & 3.29(0.93) & 9.50(1.70) & 16.3(2.55) & ---        \\ 
A1367           & n & 1.07(0.61) & 3.35(1.22) & 5.42(1.76) & ---        \\ 
A1656           & n & 1.23(0.43) & 2.27(0.62) & 3.31(0.81) & ---        \\ 
A2256           & n & 2.76(0.74) & 7.59(1.30) & 12.2(1.84) & ---        \\ 
A2319           & n & 4.84(1.61) & 11.6(3.48) & 19.9(5.12) & ---        \\ 
A3266           & n & 1.75(0.80) & 5.75(1.37) & 10.1(2.00) & ---        \\ 
A3376           & n & 0.32(0.18) & 0.79(0.41) & 1.55(0.57) & ---        \\ 
A3627           & n & 1.29(0.48) & 2.49(0.69) & 3.69(0.90) & ---        \\ 
Triang.Austr.   & n & 4.34(1.81) & 8.62(3.21) & 13.3(4.48) & ---        \\ 
%\multicolumn{}{l}{{$^{(\dagger)}$}These values are extrapolated and should
%be considered as upper limits (see text for details)} \\
\multicolumn{6}{l}{{$^{(\dagger)}$}These values are extrapolated and should
be considered as upper limits.}
\end{tabular}

\end{center}
\end{table*}
%----------------------------------------------------------------------

The global abundance  obtained by fitting simultaneously  the profiles
of both  CC  and   NCC  clusters  is $0.55\pm  0.01$.   However,  when
excluding from  the   fit the  central regions of    CC clusters (i.e.
$r<0.2~r_{200}$), this global value decreases to  $0.34\pm 0.02$.  The
higher value  obtained in the  first  case is  due to the  presence of
abundance and  surface brightness peaks at the  centre of  CC clusters
and to the emission-weighted nature of the abundance measure.  We will
show in Sect. 4 that the abundance peaks  in CC clusters are related
to local enrichment processes connected with the presence of a massive
galaxy.  We therefore assume  that the global abundance  of the ICM is
the one computed by excluding the centres  of CC clusters.  One should
note that this  value is $\sim 1/3$   of the solar abundance  when the
solar reference system  is that  given by  Grevesse \& Sauval  (1998),
corresponding to $0.23\pm  0.02$  in  the  Anders \&   Grevesse (1989)
reference system.  

\section{The ICM iron Mass}
The iron  mass enclosed within a sphere  of radius  $R$ is obtained by
radially integrating  $\int_0^R   \rho_{\rm  Fe}(r)    dV(r)$,   where
$\rho_{\rm Fe}$ is the iron density by mass  and $dV$ is the volume of
the  considered element, both measured   at  a distance $r$ from   the
centre.  Since  the deprojected iron abundance   is defined as $Z_{\rm
Fe}=n_{\rm Fe}/n_{\rm H}$ (in  units of $Z_\odot$,  that is  the solar
abundance of iron), where  $n_{\rm Fe}$ and $n_{\rm  H}$ are  the iron
and hydrogen densities (by  number) respectively, the total  iron mass
in solar units can be written as:
\begin{equation}  
M_{\rm Fe} (<R) = 4\pi A_{\rm Fe} m_{\rm H} {Z_\odot
\over M_\odot}~ \int_0^R Z_{\rm Fe}(r) ~n_{\rm H}(r)~ r^2 dr , 
\label{eq:mfe}  
\end{equation}  
where $A_{\rm Fe}$ is the atomic weight of iron and $m_{\rm H}$ is the
atomic unit mass.  To  integrate the observed  profiles at any radius,
we smoothed the metallicity  profiles applying a Savitzky-Golay filter
(Press et al.  1992, Sect.  14.8).  In Table~\ref{tab1} we present the
ICM iron masses  within overdensities $\Delta=2500$, $1000$ and $500$,
where $\Delta$ is the overdensity defined with respect to the critical
density, $\rho_{\rm c, z} = (3 H_z^2) / (8 \pi  G)$, and $r_\Delta$ is
the associated  radius.  We recall that,  at  $\Delta=2500$, 18 galaxy
clusters in  our  sample  have  detectable X-ray emission,   while, at
$\Delta=1000$($500$),  11(2) objects  are observable  (see  Fig.  1 in
EDM02).   To  estimate the ICM   iron  mass enclosed   within a  given
$\Delta$ for each cluster we  interpolate or extrapolate linearly  the
cumulative iron mass profile   up to $r_\Delta$.  Since the  abundance
and gas density profiles  are likely to remain  constant or decline at
large radii the extrapolated  values, especially those for overdensity
500, must be considered as upper limits of the iron mass.

Eq.~(\ref{eq:mfe}) is also used to estimate the iron mass excess in CC
clusters, $M_{\rm Fe}^{\rm exc}$,  namely the iron mass associated  to
the abundance peaks in the central cooling  regions of these clusters.
In  this case  we compute  the  deprojected abundance  excess, $Z_{\rm
Fe}^{\rm  exc}(r)   = Z_{\rm  Fe}(r)  - Z_{\rm  Fe}^{\rm  out}$, where
$Z_{\rm Fe}^{\rm  out}$ is the measured  average metal abundance of CC
clusters at  radii larger than  0.2~$r_{200}$,  and substitute  it  to
$Z(r)$ in  eq.~(\ref{eq:mfe}).  The  derived   iron mass excesses  for
$Z_{\rm Fe}^{\rm  out} = 0.40\pm 0.03$  (solar  units) are reported in
Table~\ref{tab1}, the  analysis  of the iron   mass  excesses will  be
discussed in Sect. 4.

%----------------------------------------------------------------------
\begin{figure}
\begin{center}
\epsfig{figure=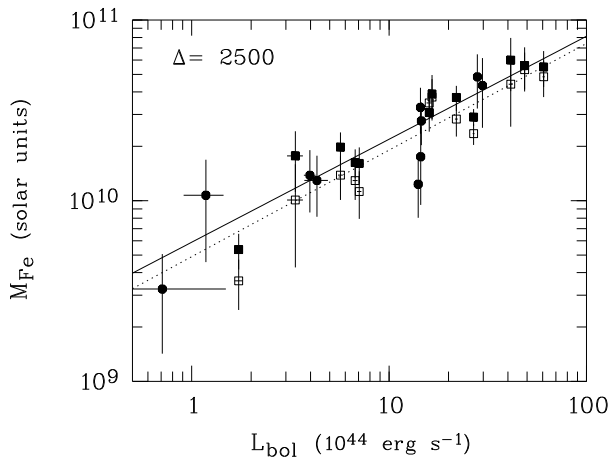,width=0.35\textwidth,angle=-90}  
\caption{
ICM iron mass as a function of the cluster X-ray bolometric luminosity
both computed within overdensity $\Delta=2500$ .  {\it Filled squares}
represent CC  clusters, whereas {\it filled  circles} are NCC objects.
{\it Open squares} show ICM iron mass of CC clusters after subtraction
of the iron  mass excess.  The {\it  solid} and the {\it dotted} lines
are  the best-fitting power law  models  (see Table~\ref{bces}) to all
clusters taking into account the total ICM iron mass and the iron mass
without the iron excess, respectively. }
\label{Fig3} 
\end{center} 
\end{figure}
%----------------------------------------------------------------------

We have investigated the relation between the  ICM iron mass and other
important observables,  i.e.  the  cluster X-ray bolometric luminosity
and temperature.  The two correlations  for $\Delta=2500$ are shown in
Fig.~\ref{Fig3} and Fig.~\ref{Fig5}, respectively.  In both figures CC
clusters are shown as filled squares,  whereas NCC systems are plotted
as filled circles.  For CC clusters we have also  plotted the ICM iron
masses without the iron mass excess associated  to the cooling regions
(open squares in both figures): $M^{\rm noexc}_{\rm Fe} = M_{\rm Fe} -
M_{\rm Fe}^{\rm exc}$.

We have characterised the dependence of the physical quantities in the
$M_{\rm Fe}-L_{\rm  bol}$ and   $M_{\rm Fe}-T_{\rm ew}$   relations by
means of a linear regression using  the algorithm described in Akritas
\&  Bershady (1996, BCES method hereafter),  which  takes into account
measurements errors and  intrinsic scatter on the  data  along both X-
and Y-axis. The linear  regression has been  applied to the logarithms
of    the  physical quantities.  In  Table~\ref{bces}    we report the
best-fit results of the bisector modification  of BCES (see Akritas \&
Bershady 1996 for details). The errors  on the best-fit parameters are
obtained from 10000 bootstrap resamplings.

%----------------------------------------------------------------------
\begin{figure}
\begin{center}
\epsfig{figure=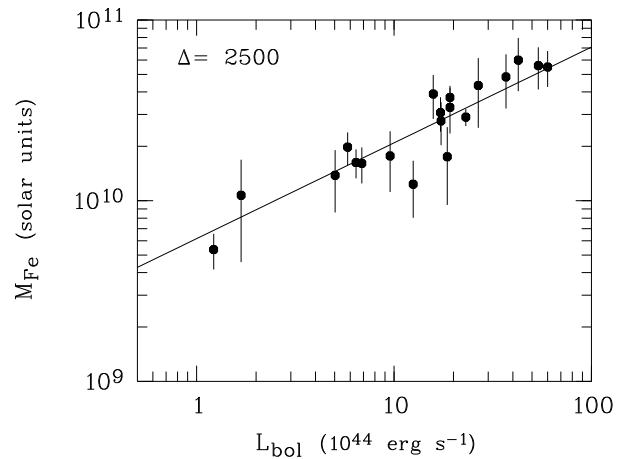,width=0.35\textwidth,angle=-90}  
\caption{ICM iron mass enclosed within overdensity $\Delta=2500$ 
   plotted as a function  of cluster X-ray bolometric  luminosity from
   the catalogue of David et  al. (1993). The  {\it solid} line is the
   best-fitting power law model (see Table~\ref{bces}).}
\label{Fig4} 
\end{center} 
\end{figure}
%----------------------------------------------------------------------

We find  that more luminous clusters posses  larger amounts of iron in
their    ICM.    The $M_{\rm  Fe}-L_{\rm   bol}$   relation,  shown in
Fig.~\ref{Fig3},    appears tight,  namely     the scatter around  the
best-fitting power  law along the y-axis ($M_{\rm  Fe}$) is small (see
Table~\ref{bces}).  The X-ray  luminosity  and the  ICM iron mass  are
both  derived using the emission  integral measured from the same {\it
BeppoSAX}  datasets and  are  therefore  subject   to some degree   of
correlation.  To test whether the $M_{\rm Fe}-L_{\rm bol}$ relation is
affected    by systematic effects   we  have further investigated this
relation using  bolometric luminosities  derived from another  totally
independent dataset, namely luminosities taken  from the work of David
et al.  (1993) (Fig.~\ref{Fig4}).  We find that the difference between
this   last  correlation  and the   one  measured  directly from  {\it
BeppoSAX} data is negligible (see  BCES results in  Table~\ref{bces}),
confirming that the $M_{\rm Fe}-L_{\rm bol}$ relation is robust.

%----------------------------------------------------------------------
\begin{figure}
\begin{center}
\epsfig{figure=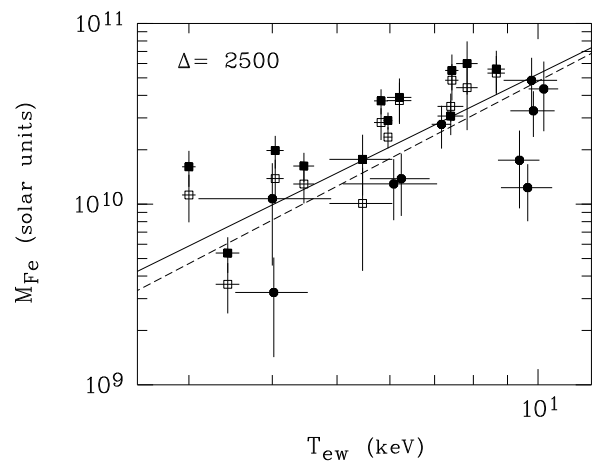,width=0.35\textwidth,angle=-90}  
\caption{ 
ICM   iron   mass  as   a   function   of   the  cluster   deprojected
emission-weighted     temperature both   computed  within  overdensity
$\Delta=2500$ .    Symbols and  lines have   the same  meaning   as in
Fig.~\ref{Fig3}.}
\label{Fig5} 
\end{center} 
\end{figure}
%----------------------------------------------------------------------

%----------------------------------------------------------------------
\begin{figure}
\begin{center}
\epsfig{figure=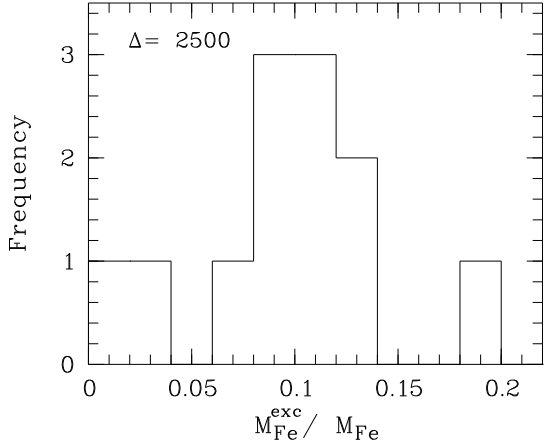,width=0.35\textwidth,angle=-90}  
\caption{Histogram of the ratio between ICM excess and ICM total iron 
mass computed at overdensity $\Delta=2500$ for CC clusters.}
\label{Fig6} 
\end{center} 
\end{figure}
%----------------------------------------------------------------------

%----------------------------------------------------------------------
\begin{figure}  
\begin{center}
\epsfig{figure=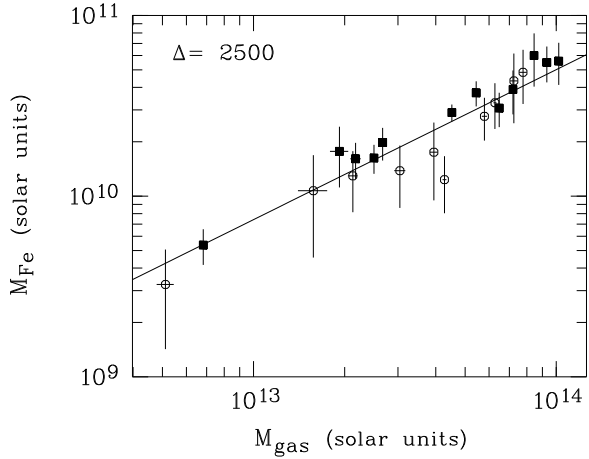,width=0.35\textwidth,angle=-90}   
\caption{ICM iron mass as a function of ICM gas mass at overdensity 2500. 
{\it Filled squares}  and {\it open circles} are  CC and NCC clusters,
respectively. The {\it solid} line is the best-fitting power law model
to all clusters.}
\label{Fig7} 
\end{center}
\end{figure}  
%----------------------------------------------------------------------

The   ICM    iron   mass  also     correlates   with   the deprojected
emission-weighted temperature.  In Fig.~\ref{Fig5} we show that hotter
clusters contain larger amounts of iron than cooler systems.  Contrary
to the $M_{\rm Fe}-L_{\rm bol}$ relation,  the $M_{\rm Fe}-T_{\rm ew}$
relation shows  a noticeable scatter  along the y-axis ($M_{\rm Fe}$),
more than a factor 2 larger than that  of the $M_{\rm Fe}-L_{\rm bol}$
relation, and  shows a clear segregation between  CC and NCC clusters.
This segregation cannot be  entirely due  to  the contribution of  the
iron excess observed in  CCs to the total  amount of iron.  Indeed, as
shown in Fig.~\ref{Fig6}, the iron mass excess is a modest fraction of
the  total   ICM  iron mass  (i.e.    only  $< 20\%$   at  overdensity
$\Delta=2500$).   We  hypothesise that  the  ICM iron  mass correlates
directly with the  cluster luminosity, whereas the correlation between
iron   mass  and  temperature  is  not  direct but  derives  from  the
combination  of  the $M_{\rm Fe}-L_{\rm  bol}$   relation and the well
known relation between  X-ray luminosities and gas temperatures (EDM02
and references  therein).  The $L_{\rm  bol}-T_{\rm ew}$ relation (see
EDM02 for   an extensive discussion  about  the luminosity-temperature
relation derived at   various  overdensities from the {\it   BeppoSAX}
sample used  in  this work; and   also Markevitch 1998 and  references
therein for a discussion of  the intrinsic scatter in this  relation),
exhibits  a  large scatter  mostly due  to  the effect  of  strong gas
cooling  in  the  cores of   CC clusters,   which   are biasing global
temperature  and luminosity measurements (Fabian   et al.  1994).  For
instance, EDM02 have shown that the scatter in the $L_{\rm bol}-T_{\rm
ew}$ relation  is reduced   by   $20\%$ when   only  CC  clusters  are
considered.

%----------------------------------------------------------------------
\begin{table*}  
\begin{center}   
\caption{Results of the best-fit analysis.  
To derive  the slope of  a relation we apply  the linear BCES bisector
estimator to the logarithm of the power law $Y = a X^b$, $\log Y = A
+B \log  X$ (i.e.  $a = 10^A,  b=B$; errors are given in parentheses).
The emission-weighted temperature, $T_{\rm  ew}$, is in units  of keV;
the luminosity, $L_{\rm bol}$, is in  $10^{44}$ erg s$^{-1}$; the iron
mass,  $M_{\rm Fe}$, and  the iron mass  after subtraction of the iron
mass excess (for  CC clusters only),  $M_{\rm Fe}^{\rm noexc}$, are in
$10^{10}  M_{\odot}$,  the gas  mass, $M_{\rm   gas}$, is  in $10^{13}
M_{\odot}$.  The scatter on $Y$ is measured as $\left[\sum_{j=1,N}
\left(\log Y_j -A -B \log X_j \right)^2 /N \right]^{1/2}$ (the scatter  
along the X-axis can be  estimated as $\sigma_{\log X}=\sigma_{\log Y}
/ B$).}
\label{bces}  
\begin{tabular}{l@{\hspace{.8em}} c@{\hspace{.8em}} c@{\hspace{.7em}}
 c@{\hspace{.7em}} c@{\hspace{.7em}} c@{\hspace{.7em}}
 c@{\hspace{.7em}} c@{\hspace{.7em}}c@{\hspace{.7em}} }
 Relation & & \multicolumn{3}{c}{ $\Delta=$2500 } & & 
 \multicolumn{3}{c}{ $\Delta=$1000 } \\
% best-fit use bces_regress_src.exe (bces.inp), 
% scatter use scatter.exe (scatter.inp), 
% expected values for MFe-Lbol use dsigma.exe
& & $A$ & $B$ & $\sigma_{\log Y}$ & & $A$ & $B$ & $\sigma_{\log Y}$\\ 
 ${\rm M_{Fe,10}-L_{bol,44}}$ 
&    &  -0.23 (0.09) & 0.57 (0.07)  & 0.13 &  & -0.23 (0.10) & 0.76 (0.07) & 0.16 \\ 
 ${\rm M_{Fe,10}^{noexc}-L_{bol,44}}$    	       				 
&    &  -0.31 (0.10) & 0.59 (0.08)  & 0.14 &  & -0.34 (0.10) & 0.80 (0.08) & 0.16 \\ 
 ${\rm M_{Fe,10}-L_{bol,44}^{(David~ et~ al.~ 1993)}}$ 	       			   
&    &  -0.21 (0.08) & 0.53 (0.06)  & 0.11 &  & -0.01 (0.12) & 0.64 (0.09) & 0.14 \\ 
 ${\rm M_{Fe,10}-L_{44}^{(0.1-2.4~ keV)}}$ 	       			  	 
&    &  -0.31 (0.11) & 0.67 (0.09)  & 0.14 &  & -0.07 (0.13) & 0.75 (0.10) & 0.19 \\ 
% ${\rm M_{Fe,10}^{noexc}-L_{44}^{(0.1-2.4~ keV)}}$ 	       			   
%&    &  -0.40 (0.12) & 0.70 (0.11)  & 0.16 &  & -0.17 (0.15) & 0.80 (0.12) & 0.23 \\ 
 ${\rm M_{Fe,10}-T_{ew}}$  			       			 	 
&    &  -1.10 (0.34) & 1.82 (0.40)  & 0.26 &  & -0.20 (0.71) & 1.15 (0.95) & 0.37 \\ 
 ${\rm M_{Fe,10}^{noexc}-T_{ew}}$ 		       				 
&    &  -1.25 (0.31) & 1.93 (0.37)  & 0.24 &  & -0.33 (0.73) & 1.25 (0.96) & 0.39 \\ 
 ${\rm L_{bol,44}-T_{ew}}$       	     					   
&    &  -1.54 (0.51)  & 3.21 (0.60) & 0.39 &  & -0.14 (0.65) & 1.78 (0.82) & 0.46 \\ 
 ${\rm M_{Fe,10}-M_{gas,13}}$ 	     	 				 
&    &  -0.13 (0.04) & 0.83 (0.06)  & 0.10 &  & -0.12 (0.06) & 0.88 (0.05) & 0.11 \\ 
% ${\rm M_{Fe,10}^{noexc}-M_{gas,13}}$    	       			 	 
%&    &  -0.21 (0.03) & 0.86 (0.05)  & 0.08 &  & -0.22 (0.05) & 0.93 (0.04) & 0.11 \\ 
\end{tabular}
  
\end{center}  
\end{table*}  
%----------------------------------------------------------------------

To test our hypothesis we   have computed the expected parameters  and
the scatter along the y-axis for  the $M_{\rm Fe}-T_{\rm ew}$ relation
by combining the best-fitting parameters  and scatters of the  $M_{\rm
Fe}-L_{\rm  bol}$ and $L_{\rm bol}-T_{\rm ew}$  relations, and we have
compared them with  those derived from direct  fitting of  the $M_{\rm
Fe}-T_{\rm ew}$  relation.  We find that  the parameters computed from
the $M_{\rm  Fe}-L_{\rm bol}$ and   $L_{\rm bol}-T_{\rm ew}$ relations
($A=-1.11,B=1.83$ for $\Delta=2500$), are in good agreement with those
derived  from direct fitting of  the  $M_{\rm Fe}-T_{\rm ew}$ relation
(see Table~\ref{bces}).  Most  importantly,  we find that  the scatter
along  the  y-axis ($M_{\rm Fe}$) observed   in the $M_{\rm Fe}-T_{\rm
ew}$ relation ($\sigma_{\rm log  M_{Fe}}=0.26$) is consistent with the
one derived  by combining  the  $M_{\rm  Fe}-L_{\rm bol}$ and  $L_{\rm
bol}-T_{\rm   ew}$ relations   ($\sigma_{\rm  log M_{Fe}}=0.27$).   We
therefore conclude that the large scatter and the segregation observed
in    the $M_{\rm Fe}-T_{\rm  ew}$ distribution   is  due to the large
dispersion in X-ray luminosities for a given temperature, and that the
most direct relation  is  the one  between the ICM  iron mass  and the
X-ray luminosity.

In Fig.~\ref{Fig7} we  explore the relation between  the iron mass and
the gas mass  in the  ICM.  Such a  relation  is expected as a  direct
consequence of the $M_{\rm  Fe}-L_{\rm bol}$ relation since  the X-ray
luminosity   is  related  to  the  square of  the  gas  mass  ($L_{\rm
bol}\propto   n_H^2  T^{1/2}$).  We  find that   the  iron mass scales
linearly with the gas mass (see Table~\ref{bces}), which is equivalent
to  saying that  all nearby clusters   have  the same iron  abundance,
$Z_{\rm Fe}\propto M_{\rm Fe}/M_{\rm gas}$.  Since the iron in the ICM
has  been formed in  stars this  result supports a  scenario where the
mass in stars  in clusters is closely  related to the intracluster gas
mass.  At   overdensity  $\Delta=2500$ the  iron to  gas   mass ratio,
$M_{\rm  Fe}/M_{\rm  gas}$, ranges   between $2.9-9.2\times  10^{-4}$,
which corresponds to an iron abundance in  solar units of $0.22-0.71$.
We  note  that   this measure    of  the   ICM  metallicity is     not
emission-weighted  as it  is  derived from  direct integration  of the
deprojected iron   abundance  and     gas density  profiles.      This
mass-weighted   metallicity    is   $\sim  15\%$    smaller  than  the
emission-weighted projected abundance  measured directly from the data
within the same overdensity,  $\Delta=2500$.  The difference is due to
the abundance and surface brightness peaks in CC clusters.

%----------------------------------------------------------------------
\begin{figure}
\begin{center}
\epsfig{figure=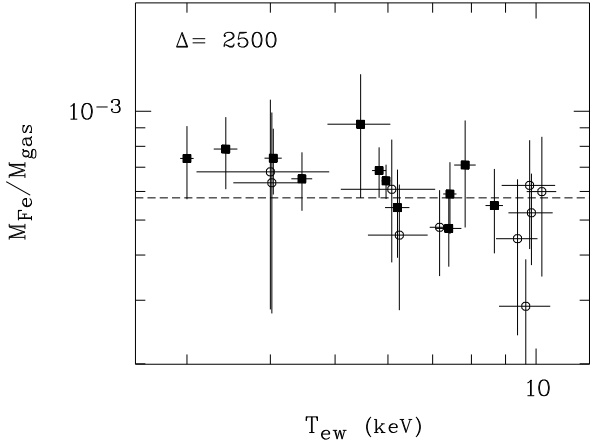,width=0.35\textwidth,angle=-90}  
\caption{Ratio between ICM iron mass and gas mass, i.e. metallicity, 
as a function  of temperature.  {\it Filled  squares}  are CC clusters
and {\it open circles}  are NCC systems.  The {\it dotted} line is the
best-fitting constant model to all clusters.}
\label{Fig8} 
\end{center} 
\end{figure}
%----------------------------------------------------------------------

Finally, as  illustrated in Fig.~\ref{Fig8} we  show that  the ratio
between the iron  and gas mass  (at $\Delta=2500$)  does  not show any
trend with the cluster temperature.  A constant  model fitted over the
distribution of  all  clusters gives a  mean   metallicity of $0.44\pm
0.02$  ($M_{\rm Fe}/M_{\rm  gas}=  5.76\pm 0.4  \times 10^{-4}$,  with
$\chi^2$/dof=18.2/21). Since CC span  a different range in temperature
with respect to  NCC systems  we have fitted  the  two cluster classes
independently:  the best-fitting constant for  CC clusters is $6.3 \pm
0.4 \times 10^{-4}$ with $\chi^2$/dof=6/11), and that for NCC clusters
is $4.5 \pm 0.6\times  10^{-4}$ with  $\chi^2$/dof=5/9), corresponding
to iron abundances of $0.51\pm 0.03$ and $0.36\pm 0.05$, respectively.
We   find  that the CC cluster   distribution   shows a  mild negative
gradient with increasing temperature: indeed, by performing a fit with
a  constant plus  a linear component   we find a statistically  modest
improvement according to the $F$-test (Prob$_{\rm F-test} = 0.95$).

\subsection {The local ICM iron Mass Function and the Metal Content in 
Clusters}
The $M_{\rm Fe}-L_{\rm bol}$ relation allows us to estimate a reliable
ICM iron mass once  the cluster X-ray  luminosity is known.  From this
relation it is straightforward to derive the iron mass function of the
ICM (X$M_{\rm Fe}$F)  in  the local   Universe using the   local X-ray
luminosity function of clusters (XLF).

We have computed the X$M_{\rm Fe}$F by assuming  the XLF of the REFLEX
sample (B\"ohringer et al.  2002,   their eq.  4)  in  the case of  an
$\Omega_m=0.3$  and  $\Omega=0.7$  Universe  (with parameters   of the
Schechter   function  $n_0=1.07\times     10^{-7}$         Mpc$^{-3}$,
$\alpha=1.69\pm 0.045$,   $L_*=8.36^{+0.9}_{-0.8}     10^{44}$    ergs
s$^{-1}$), and  our  scaling  relation  $M_{\rm Fe}-L_{(0.1-2.4~keV)}$
computed with luminosities converted  into the 0.1-2.4 keV energy band
(parameters A=-0.31,  B=0.67, see Table~\ref{bces}).  To  estimate the
clusters contribution to the iron mass budget of the local Universe we
have integrated the X$M_{\rm Fe}$F over a  given iron mass range.  The
integral limits  have  been chosen  from  the $M_{\rm Fe}-T_{\rm  ew}$
relation  for a  minimum and  maximum  temperature of   3 and 10  keV,
respectively.  The resulting iron mass in clusters with respect to the
total critical density, $\Omega_{Fe}=\rho_{Fe}/\rho_c$, is $4.54\times
10^{-8}$.

From this we have estimated the total metal budget in nearby clusters,
by  summing over all relevant  metals considering a Grevesse \& Sauval
(1998) solar abundance   and  assuming that the enrichment   is mostly
given by SNII  (we have assumed  metal-to-iron mass ratios accordingly
with Nomoto  et al.  1997).   We compute  an upper limit  of the metal
content in  clusters relative to  the critical density, $\Omega_Z$, of
$\sim 1.61\times 10^{-6}$,  by excluding only the  masses of H  and He
from the calculation.  By excluding the masses of C and N too, as they
are  both metals not  yet measured in  the ICM (see  Finoguenov et al.
2003),  we  obtain the  lower    limit of  $\Omega_Z  \sim  1.47\times
10^{-6}$.   This  contribution is about  a  factor 3 smaller than that
estimated  independently, with  different  assumption and  methods, by
Finoguenov et al.  (2003).

\subsection {The cluster IMLR}
The concept  of ICM iron mass  to light ratio  ($M_{\rm Fe}/L_B$, IMLR
hereafter)  as the ratio  of the total iron mass  in the  ICM over the
total optical luminosity of    galaxies   in the cluster  was    first
introduced by Ciotti  et al.  (1991).  This  is a fundamental quantity
to  understand the  ICM enrichment   (Renzini 1997).  Renzini et   al.
(1993)  derived   a typical value  of  the  IMLR for rich  clusters of
$0.01-0.02$ ($M_\odot/L_\odot$).  These values were obtained computing
the iron  mass in the  ICM as the product of  the iron abundance times
the  mass of the ICM gas:  $M_{\rm Fe}\propto Z_{\rm Fe} M_{\rm gas}$,
using the global cluster value for the abundance, $Z_{\rm Fe}$.  Since
rich relaxed clusters show abundance gradients,  this way of measuring
the IMLR tends to overestimate the total amount of iron present in the
ICM (see discussion in Sect. 2).

%--------------------------------- ------------------------------------
 \begin{figure}
 \begin{center}
 \epsfig{figure=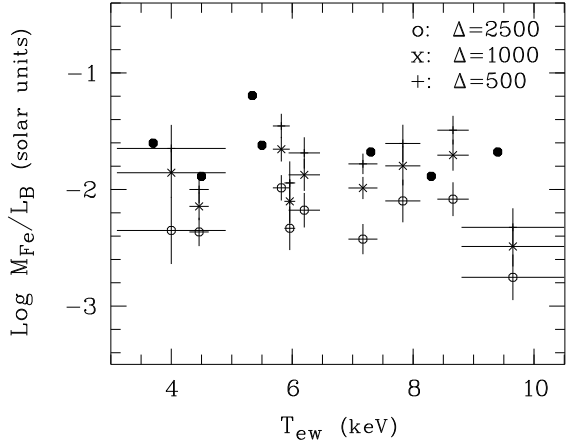,width=0.35\textwidth,angle=-90} 
 \caption{ICM IMLR, $M_{\rm Fe}/L_{\rm  B}$, as a function  of cluster
 temperature at $\Delta=2500$  of clusters A1367, A2199,  A1795, A426,
 A85, A2256,  A2029,    A2142   and A1656  (ordered     by  increasing
 temperature).    {\it  Filled circles}  are   IMLR  values taken from
 Renzini (1997;  the 7 cluster in common  with  our sample), {\it open
 circles}, {\it crosses} (x) and {\it  crosses} (+) are IMLR estimated
 using iron  masses  from  this work computed  within
 $\Delta=2500$, 1000 and 500, respectively.}
 \label{Fig9} 
 \end{center}
 \end{figure}
%----------------------------------------------------------------------

We  re-compute   the IMLR using    our  ICM iron  masses estimated  by
integrating radially $Z_{\rm  Fe}$ with eq.~(\ref{eq:mfe}).  For the 9
clusters  with available optical B-band   luminosity (we use the  same
$L_{\rm B}$ used by  Renzini 1997 and Arnaud et  al.  1992), the  mean
value  of the IMLR at the  various  overdensities is $5.8\pm 0.9\times
10^{-3}$ ($\Delta=2500$),  $1.3\pm  0.2  \times 10^{-2}$  ($\Delta   =
1000$) and $2.0 \pm 0.3 \times 10^{-2}$ ($\Delta=500$).
When  the iron mass excess is  subtracted from the contribution of the
CC clusters the IMLR  at overdensity 2500  becomes $5.0\pm  0.7 \times
10^{-3}$, therefore this exclusion has a negligible effect.

By comparing  our measurements with  those of Renzini (1997),  we find
that their values are in excess of ours by about a factor 2 when using
our $\Delta=1000$  values and by  about  a factor  1.3  when using our
$\Delta=500$ values.  In a recent  review, Renzini (2003), has revised
his values by reducing  them by $\sim 30\%$, as  a consequence of this
change his and ours estimates are now in reasonable agreement.

In Fig.~\ref{Fig9} we plot  our IMLR and those of  Renzini (1997) as a
function of the cluster temperature.  We do not find any dependence of
the IMLR with the temperature in  agreement with what previously found
by Renzini (1997).

\section{The iron Mass Excess in Cool Core clusters}
NCC clusters have  flat  abundance profiles, whereas CC  clusters have
abundance excesses  in  their cores (Fig.~\ref{Fig1}).   In this Sect.
we address the question of assessing the cause of this difference.  In
our previous work  (DM01)  we compared  the observed  abundance excess
profiles of four clusters (A2029,   A85,  A496 and Perseus) with   the
abundance excess profiles  expected when the metal excess distribution
traces the light distribution  of early-type galaxies of  the cluster.
We  found a good agreement  between the observed and expected profiles
and,  more interestingly, we  found that  the  excess in  the expected
profiles is entirely due to the light profile of the brightest cluster
galaxy  (BCG).  This  indicates that the  abundance  excess could have
been producted by the BCG itself through direct ejection of metal rich
gas  via SN- or   AGN- induced winds.   A  similar conclusion has been
reached  by  other  authors (Fukazawa et   al.    2000 and  references
therein), which suggested that the central  galaxy is the cause of the
observed abundance peaks.  This  is    furthermore supported by    the
observational evidence that the X-ray emission peak  of CC clusters is
{\it always}   centred on  the BCG (e.g.    Schombert 1988,   Peres et
al. 1998, Lazzati \& Chincarini 1998).

To explore  in greater detail the possibility  that the BCG  itself is
producing the observed  iron  abundance  peak we have   estimated  the
amount of iron that  such a galaxy is   able to eject during its  life
performing the exercise reported in the  following.  We have collected
from the literature optical Kron-Cousin $R_{c}$-band magnitudes of the
BCGs  integrated through an   aperture of constant  physical radius of
$\approx 18$ kpc  (Postman \& Lauer 1995).   For A85,  A496, A1795 and
A2029 we have converted $r$-gunn magnitudes on the same aperture taken
from Hoessel, Gunn \&  Thuan (1980) into   $R_c$ magnitudes using  the
relation $r-Rc=0.37$ given in Fukugita  et al.  (1995), while  optical
magnitudes   were   not available  for  the   central  galaxies  of 2A
0335$+$096   and PKS   0745$-$191.     Typical errors of  the  adopted
magnitudes are less   than 0.05 mag,  while redshifts  of galaxies are
given within an error of 0.0002. All the apparent magnitudes have been
corrected for the galactic extinction assuming the E(B-V) values given
by Burstein and Heiles (1984).  Then, the stellar  mass content of the
BCGs has been derived from their  absolute $R_{c}$ magnitudes and from
the mass-to-$R_{c}$ band light ratio.  The adopted ranges of values of
k-corrections and mass-to-light ratios have  been calculated using the
latest version of the stellar population  synthesis code of Bruzual \&
Charlot  (1993), considering two   possible models:  a Simple  Stellar
Population  (SSP) and an  exponentially declining  star formation rate
with  a timescale $\tau$ = 1.0  Gyr.  Both population synthesis models
have been  built assuming a Salpeter  Initial Mass Function  (IMF) and
solar  metallicity  (Z=0.020), and  a typical age   of 13 Gyr has been
chosen to  represent   our massive  ellipticals.   We obtain   typical
mass-to-light  ratios  of 8.6 and  7.9 (solar  units) for the  SSP and
$\tau = 1.0$  Gyr models respectively (assuming M$_{\odot,R_{c}}=4.4$,
derived  from Allen 1973).  It  is worth noting that  the  mass in the
above ratios  includes the total  mass  of gas  involved in  the  star
forming   process.    Indeed,  considering   the stellar   mass really
``locked'' into  stars (i.e.  the  difference from the total processed
gas   and the  processed gas  returned   to the interstellar medium of
galaxies) we would obtain $30\%$ lower  masses.  Variations of the age
of the models in the range of 10 - 15 Gyr would produce differences in
the  corresponding mass-to-light ratios  of less than $20\%$ while the
major source of uncertainty in  the mass estimate  of the BCGs stellar
content comes from the choice of IMF.  Indeed, assuming a Scalo IMF we
would   obtain stellar  masses  lower  by  a  factor  of  2,   while a
Miller-Scalo IMF would bring stellar masses down  by more than $ 30\%$
with  respect to those obtained with  the  chosen Salpeter IMF.  Given
the low redshift of the BCGs in the selected sample, k-corrections are
small (i.e., lower than 0.1 mag) and their change due to variations of
the models parameters do not affect the  final estimate of the stellar
masses significantly.    Putting  everything together   the  resulting
stellar mass content of   the BCGs in the   selected sample is in  the
range $\sim 6\times 10^{11} - 4\times 10^{12}$ $M_\odot$.

We  have then converted  the stellar mass  range  into a range of iron
masses ejected into the ICM by these galaxies considering the chemical
evolution of ellipticals as modelled by Pipino et al.  (2002) (we have
interpolated the  values given  in their  Table 4).   The ejected iron
mass range obtained  is $\sim 1-5\times  10^9$ $M_\odot$.  It is worth
noting that  these masses are lower  limits to  the iron mass ejection
since  the  optical  magnitudes within 18  kpc  do  not take into full
account the large stellar halos which are  present in many of the BCGs
(e.g., Kemp \& Meaburn 1991 found  a 0.6 Mpc  halo in the cD galaxy of
A3571).

We have  compared this  expected iron mass   range with the  iron mass
excess   measured in the  core  of  CC  clusters  (see  Sect.  3 and
eq.~\ref{eq:mfe}).   Since    the  observed   values  range    between
$0.5-9.5\times 10^9$ $M_\odot$ (Table~\ref{tab1}) and considering that
the estimate is characterised by the uncertainties described above, we
conclude that the  BCGs  are able to   produce the observed iron  mass
excess during their lifetime.

%----------------------------------------------------------------------
\begin{figure}
\begin{center}  
\epsfig{figure=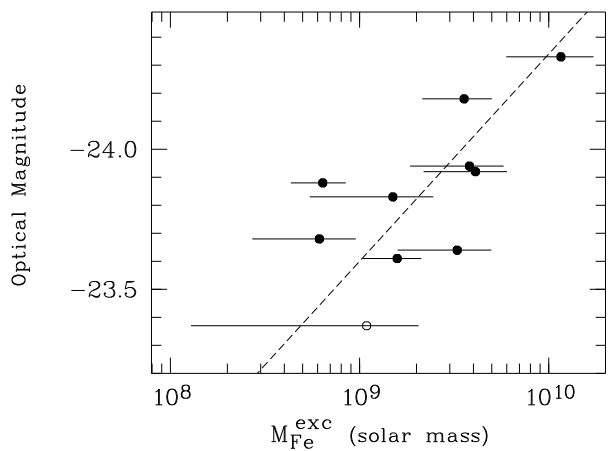,width=0.35\textwidth,angle=-90}  
\caption{BCG optical magnitude as a function of iron mass excess.
{\it Open circle} marks the binary cluster A2142.   Dashed line is the
best-fitting power law excluding A2142.}
\label{Fig10}
\end{center} 
\end{figure}  
%----------------------------------------------------------------------

If we assume that the  iron  mass excess  measured  in the core of  CC
clusters is  produced   by the  BCG   alone, then we  expect  that the
properties of  the BCGs, more  specifically their  optical magnitudes,
which are a measure  of their stellar  masses, correlate with the iron
mass excess.  Indeed this correlation seems  to be present in our data
as  illustrated in Fig.~\ref{Fig10}.  The cluster  marked with an open
circle in the Fig. is A2142, which has an abundance profile consistent
with  being  flat (DM01):  as  discussed in  DM01  the  cause  of this
flattening is  probably physical as  A2142 is  a known  binary  system
(e.g.  a cluster that    contains  two  bright galaxies    of  similar
magnitudes), which experienced  a  recent merger event   that probably
influenced the  abundance peak in its  core (Oegerle, Hill \& Fitchett
1995,  Buote \&  Tsai  1996, Henry \& Briel   1996, Markevitch et  al.
2000).   In  the rest of  this  work we will  exclude  A2142  from our
analysis because  it is a peculiar cluster   that cannot be considered
representative of the  CC clusters class.  Nevertheless,  this cluster
will be always  plotted in  the next figures   and marked as an   open
circle.

We find that the optical magnitude and  the iron mass excess correlate
(the  non-parametric  correlation  coefficients  for the  9   BCG with
available  optical magnitude is 0.67), and,  although there is a large
scatter in our relation and the statistics  is small, we find that the
dependence between the two quantities  is consistent with being linear
(the   best-fitting parameters are  reported in Table~\ref{bces_exc}).
If we  assume  that the BCG   is the  sole origin  of  the iron excess
observed in CC's cores and that  the metals ejected  by the BCG remain
confined in the vicinity  of the galaxy,  then this relation may imply
that the efficiency of the mechanisms that play a role in transporting
the metals from the galaxy to  the ICM may  be roughly the same in all
clusters.  A larger  sample of BCGs in  CC clusters and deeper optical
photometry,  both at optical and  NIR wavelengths, are needed to study
in greater detail this important relation.

%----------------------------------------------------------------------
\begin{table}  
\begin{center}   
\caption{Results of the BCES best-fit analysis. The parameters A, B and 
the  scatters   $\sigma_{\log X}$, have   the   same  meaning  as   in
Table~\ref{bces}.  We have considered here only quantities computed at
overdensity  2500.  The   last  column reports Spearman's   Rank-Order
correlation coefficient, $\rho$, for  a set of physical quantities; in
parentheses is reported the   probability ($P_\rho$) of  exceeding its
value under  null hypothesis of not associated  dataset (a small value
in probability indicates significant correlation).}
\label{bces_exc}  
\begin{tabular}{l@{\hspace{.8em}} r@{\hspace{.8em}} r@{\hspace{.7em}}
 r@{\hspace{.7em}} r@{\hspace{.7em}} r@{\hspace{.7em}}}
%\begin{tabular}{l@{\hspace{.8em}} r@{\hspace{.8em}} r@{\hspace{.7em}}
% r@{\hspace{.7em}} r@{\hspace{.7em}} r@{\hspace{.7em}} r@{\hspace{.7em}}}
%
% See bces.doc for documentation
% best-fit use bces_regress_src.exe (bces.inp), 
% scatter use scatter.exe (scatter.inp), 
% associa/rank table col1 col2 S
% sel/tab BCG_mass_out :name.ne."*2142*".and.sequence.lt.11
% expected values for MFe-Lbol use dsigma.exe
Relation   & $A$~~~~~~~ & $B$~~~~~~~ & $\sigma_{\log Y}$ & $\rho (P_\rho)$~~~~ \\
\\ 
 ${\rm M_{Fe,9}^{exc}-T_{ew}}$ 
& -1.25 (0.66) & 2.23 (0.83)   & 0.26~  &   0.84 (1e-3)  \\
 ${\rm M_{opt}-T_{ew}}$
& -22.3 (0.67) & -2.04 (0.86)  & 0.20~  &  -0.72 (0.03)  \\
 ${\rm M_{opt}-M_{Fe,9}^{exc}}$    
& -23.6 (0.20) & -0.74 (0.41)  & 0.22~  &  -0.67 (0.05)  \\
% ${\rm M_{Fe,9}^{exc}-L_{cool,44}}$    
%& -0.08 (0.10) & 0.51 (0.15)   & 0.26~  &   0.76 (7e-3)  \\
% ${\rm M_{opt}-L_{cool,44}}$
%&  ---~~~~~~~  &  ---~~~~~~~   & ---~~~ &  -0.45 (0.22)  \\
% ${\rm M_{Fe,9}^{exc}-M_{tot,14}}$ 
%& -0.13 (0.16) & 1.10 (0.30)   & 0.17~  &   0.88 (3e-4)  \\
\end{tabular}
  
\end{center}  
\end{table}  
%----------------------------------------------------------------------

\subsection{The Cluster/BCG connection}
At the spatial    scales examined by  {\it  BeppoSAX}  observations CC
clusters appear as dynamically relaxed objects.  X-ray observations of
their  central   regions   show in  general  simultaneously  declining
temperature profiles (e.g.   Allen  et  al.  2001), increasing    iron
abundance  profiles and  surface  brightness excesses with  respect to
single $\beta$-model profiles (e.g.   Mohr et al.  1999).   When these
properties are present, optical observations  {\it always} reveal  the
presence of  a BCG located  at the bottom  of  the cluster's potential
well, thus there must  be some physical  connection between  all these
observational facts.    In this Sect.   we   try  to  investigate  the
physical link between  the cluster and   the BCG.  We expect that  the
relation between their properties, such  as  optical magnitude of  the
galaxy and cluster temperature or iron  mass excess, may give us clues
on the way clusters and BCGs formed.

A number of   theories have been  proposed  for the formation  of  the
central  giant  galaxies  in  clusters.   The first  models, which are
sometimes  generically  labelled  as ``galactic cannibalism'', require
that a cluster (with its potential well) has already formed and indeed
these  models predate the advent of  the hierarchical scenario for the
structure formation.  These models have been found to be inadequate to
explain   the bulge  of the  BCGs  because   the dynamical  timescales
involved are too long to reproduce  the observed bulge luminosities of
these galaxies (e.g. Tremaine 1990).

Another model requires that the formation of  the BCG is significantly
influenced   by gas accretion   and  enhanced  star formation   from a
cooling-flow (Fabian 1994).  The final fate of the cooling flow gas is
to cool  down, accreting significantly    on the central  galaxy,  and
triggering star formation.  This process would require BCGs to have an
anomalously blue colours which  are not observed  (e.g. Andreon et al.
1992, McNamara \& O'Connel 1989,  1992).  However, recent works  based
on {\it  XMM-Newton}  (e.g.  Peterson  et  al.   2001, Kaastra et  al.
2001, Tamura et al.  2001, Molendi \& Pizzolato 2001, David ) and {\it
Chandra}  (e.g.  David  et al.   2001, Ettori  et  al.  2002, Blanton,
Sarazin \& McNamara 2003) observations  of clusters have shown that in
the  core   of CC clusters the   gas  does not  cool  beyond a minimum
temperature of $\sim 1-3$ keV   (Fabian et al.  2001).  Many   authors
(Churazov et al.  2002, Kaiser \& Binney 2003, Quilis, Bower \& Balogh
2001, Ruszkowski \& Begelman 2002, Br\"uggen \& Kaiser 2002, Brighenti
\& Mathews 2002, Ciotti \& Ostriker 2001) have  suggested that the gas
is prevented from  cooling by some  form of heating. We  conclude that
cooling  flows  are no   longer a viable   alternative to  explain the
formation of BCG galaxies.

The  last and most  popular model  is   galactic merging in  the early
history of the formation of the cluster.  This model predicts that the
BCG forms through mergers of several massive galaxies along a filament
early in the cluster history (Merritt  1985, Tremaine 1990).  Since in
the hierarchical scenario  smaller structures form earlier than larger
structures, a peculiarity  of this model is  that the formation of the
BCG  (or at  least  of its  bulge)   predates  or  coincides  with the
formation of the  cluster.   Numerical simulations (West  1994,
Garijo et  al.   1997, Dubinski 1998)   favour this  picture.   In the
simulation  of Dubinski (1998)   galaxy  merging naturally produces  a
massive central galaxy with surface brightness and velocity dispersion
profiles very similar to those observed in BCGs; the bulk of the BCG's
luminosity forms early in the cluster's history, after a rapid merging
of  the  most massive galaxies  and  further accretion does not change
significantly the mass   of the galaxy.   Other indications supporting
the hypothesis of hierarchical assembly of BCGs  come from large scale
observations, that  show that BCGs tend to  be aligned with the galaxy
and  X-ray gas distributions.    This  tendency continues  to  cluster
scales up  to $\sim 20-30$  h$^{-1}$ Mpc  (Melott, Chambers \&  Miller
2001) and irregular clusters  show a tendency  to be more aligned with
their neighbours   and    are preferentially  found   in  high-density
environments (Plionis \& Basilakos 2002).  Furthermore, BCGs are often
very  flat and  their  flattening  is  not primarily  due  to rotation
indicating  that  they  probably form with    a triaxial structure.  A
natural explanation for this behaviour can  be given in the context of
aspherical accretion from filaments.

%---------------------------------------------------------------------
\begin{figure}
\begin{center}
\epsfig{figure=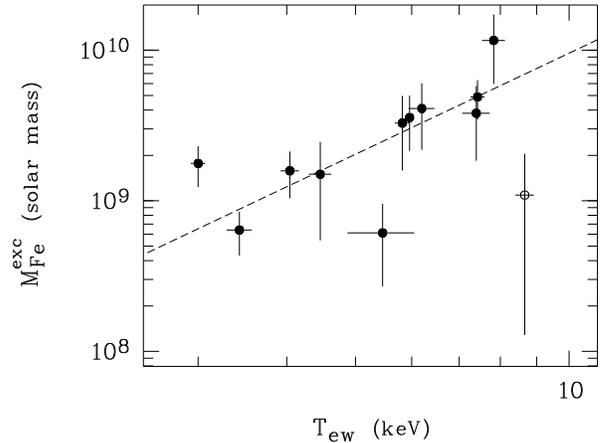,width=0.35\textwidth,angle=-90}
\caption{Iron mass  excess  in  cool core  clusters plotted  as   a
   function of the cluster temperature  at overdensity 2500. {\it Open
   circle} marks binary  cluster A2142.  The dashed  line is  the BCES
   best-fitting power law computed excluding A2142.}
\label{Fig11}
\end{center} 
\end{figure}
%----------------------------------------------------------------------

%----------------------------------------------------------------------
\begin{figure}  
\begin{center}
\epsfig{figure=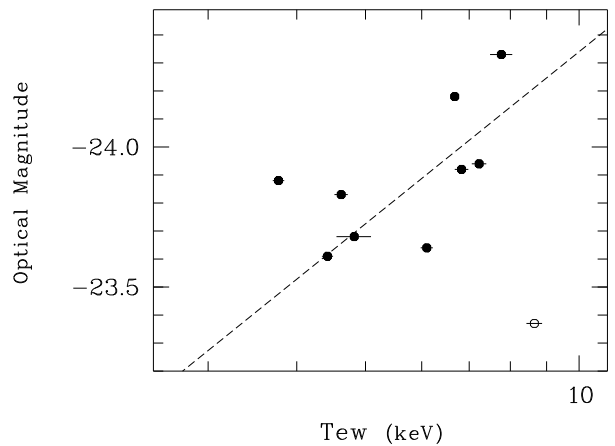,width=0.35\textwidth,angle=-90}  
\caption{BCG optical magnitude as a function of cluster temperature 
at  overdensity  2500. {\it Open   circle}  marks the  binary  cluster
A2142. Dashed line is the best-fitting power law excluding A2142.}
\label{Fig12}
\end{center} 
\end{figure}  
%----------------------------------------------------------------------

The {\it BeppoSAX} data show that the iron mass excess correlates with
the cluster  temperature (at  $\Delta=2500$)  (Fig.~\ref{Fig11}).  The
best-fitting  parameters obtained using  our eleven CC clusters (A2142
is excluded, see discussion     in   Sect.   4), are   reported     in
Table~\ref{bces_exc}.  Since the temperature is related to the cluster
gravitational   mass   ($M_{\rm  tot}\propto   T^{3/2}$),   a possible
explanation of the observed correlation  is that more massive clusters
contain more massive BCGs,   which are producing larger  quantities of
iron through feedback mechanisms  during their life.  If  this picture
is correct  than we expect that  the optical luminosities of  the BCGs
correlate both  with the cluster temperatures {\it  and} the iron mass
excesses.  We   have  already  shown  in   the   previous  Sect.  (see
Fig.\ref{Fig10}), that the  optical magnitude  of the BCG   correlates
with the  iron mass excess.   The  other expected correlation, $M_{\rm
opt}-T_{\rm ew}$, is  plotted in Fig.~\ref{Fig12}.  The non-parametric
correlation   coefficients  for the  9   BCGs  with  available optical
magnitude  is   0.72   (Prob=0.03),  indicating  the   two  quantities
correlate.   The best-fit  to our data   (A2142 excluded) is  given in
Table~\ref{bces_exc}.  This correlation implies that the BCG formation
is closely  related to that of  the whole  cluster and that subsequent
mergers  with other cluster galaxies  and other accretions events with
smaller  clusters or  groups  produce little  change   in its  overall
properties, or at least of its bulge, if this where  not the case than
the   properties  of the  BCGs   would depend  on the  different  mass
aggregation  history  of each cluster  and  we  would not  observe the
correlations reported     in  Fig.~\ref{Fig10},   Fig.~\ref{Fig11} and
Fig.~\ref{Fig12}.  A positive    correlation  between   the    optical
magnitude of the BCG and the hot  gas temperature of its host cluster,
$M_{\rm opt}-T_{\rm ew}$, has also been found by other authors (Edge
\& Steward  1991,  Edge 1991, Katayama   et al.   2003), using  larger
samples of BCGs.

%----------------------------------------------------------------------
\begin{figure}
\begin{center}  
\epsfig{figure=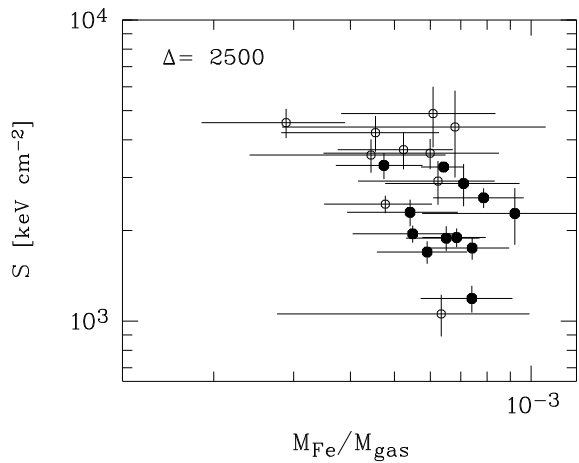,width=0.35\textwidth,angle=-90}  
\caption{Entropy versus iron abundance expressed in the form $M_{\rm 
Fe}/M_{\rm   gas}$ for CC  ({\it filled  circles}) and  NCC ({\it open
circles})  clusters.   Both  quantities  are  computed at  overdensity
2500.}
\label{Fig13}
\end {center} 
\end{figure}  
%----------------------------------------------------------------------

As discussed  above, the scenario  where the properties of the central
giant galaxy are related to  those of the  hosting cluster is expected
in  the hierarchical formation  of the  structures.  In  a recent work
based on hydrodynamic  simulations of  large-scale structure formation
including  radiative  cooling, Motl  et al.   (2003) suggest that cool
cores  in  rich  clusters form  and   grow in mass  predominantly from
hierarchical assembly of discrete subclusters, that contain pre-cooled
gas and low  temperature and  entropy  cores, and that flow  into  the
clusters along the network filaments.  Given  the larger ages of these
accreting subclumps and the  presence of a  high fraction of condensed
cool gas, it is reasonable to assume that enrichment processes through
star  formation feedback had time to  take place before accretion.  An
indications that the  gas  in  the  accreting sub-clumps   is  already
enriched before cluster relaxation may come from X-ray observations of
BCGs  in binary clusters (i.e.    clusters  with two BCGs).  The  Coma
cluster can be considered the prototype  of a binary clusters, hosting
two  dominant galaxies  (NGC 4889  and   NCG 4874).  Vikhlinin  et al.
(2001) analysing a {\it Chandra}   observation of these two   galaxies
found  that both BCGs are  able to retain  part of their  ISM and that
both  ISM have  mean abundances larger  than that  of  the surrounding
hotter ICM.   The analysis of {\it  XMM-Newton} EPIC data confirms the
{\it Chandra}    results  (Molendi  private   communication).  In  the
hierarchical model we expect that during the relaxation process of the
Coma cluster the two BCGs will eventually  merge into a unique massive
galaxy with  an associated abundance  peak and a   cool core.  In this
model    it is natural   to  expect cool  core   clusters to also show
abundance excesses in their cores   as the  results of the   accretion
process.  In other  words we expect regions  with lower entropy  (with
the entropy defined as $S = kT_e/n^{2/3}_e$ keV  cm$^{-2}$) to be also
characterised by higher metal  abundances.  This is actually  the case
as    shown in   Fig.~\ref{Fig13},    where the  entropy  measured  at
overdensity $\Delta=2500$  (about $1/3-1/4$ of  the virial radius) are
plotted  as a function  of  the iron  abundances  measured at the same
overdensity for both  CC and NCC clusters.  We  find that  CC clusters
tend to  populate a  different  region of the  entropy-abundance plane
with respect to   NCC clusters, namely  a  region where lower  entropy
values correspond to higher abundances.

\section {Summary}
In  this paper we  have derived  ICM  iron masses for  a  sample of 22
nearby rich  cluster   of galaxies, observed   with {\it BeppoSAX},  by directly
integrating the deprojected iron  abundance and gas  density profiles.
In the first  part  of the paper  we  have discussed  the global metal
content of our cluster sample while in the second we have concentrated
on the  iron mass   associated to  the  abundance excess  found  in CC
clusters.  Our main results for the first part are as follows.

\begin{enumerate}
      
\item 
The deprojected abundance profiles are strongly peaked for CC clusters
while they remain  constant  for NCC systems   in agreement with  what
previously found for the   projected  profiles published in  DM01.  

\item 
The   relationship between ICM      iron mass and    other fundamental
quantities is through the  gas mass.  Since the  iron  in the ICM  has
been formed in stars our result supports a scenario  where the mass in
stars in clusters is closely related to the ICM gas mass.

\item 
We have used the  ICM iron mass  vs.  X-ray cluster luminosity scaling
law, and the local X-ray luminosity function, to  derive the iron mass
function for the  local  universe.  By integrating this   function, we
have derived  the iron and  total metal contribution of local clusters
to the metal budget of the universe, $\Omega_{\rm Fe}$ and $\Omega_Z$,
finding them  to be   $\sim 4.5\times   10^{-8}$ and  $\sim  1.5\times
10^{-6}$, respectively.

\item 
The mean IMLR computed at various overdensities for  the 9 clusters in
our sample with available  optical luminosities range  between $5.8\pm
0.9\times 10^{-3}$  at $\Delta=2500$ and $2.0  \pm 0.3 \times 10^{-2}$
at $\Delta=500$.  These values are  consistent with the revised values
published by Renzini (2003).

\end{enumerate}

Our main results for the second part are as follows.

\begin{enumerate}

\item 
The  iron mass  associated  to abundance excess in  CC  clusters is of
$\sim 0.5-9.5\times 10^9 M_\odot$,  this is about  $10\%$ of the total
ICM iron mass  (at $\Delta=2500$).  Using population synthesis  models
to estimate the BCG stellar mass from its observed $R_c$ magnitude and
chemical enrichment  models to derive  the iron mass ejected  into the
ICM from the stellar mass, we find that the BCG is able to produce the
iron mass excess observed in CC clusters during its life.

\item 
We  find that the  optical light of  the BCG  correlates with both the
cluster temperature and the iron mass excess, implying that there is a
strong correlation between the galaxy  and the cluster potential well.
This   result  favours  current  hierarchical formation   scenarios of
structures where the BCG is assembled  at the beginning of the cluster
formation process from pre-enriched subclumps.

\end{enumerate}

In conclusion   we  would like  to stress   that more detailed optical
observations are needed to derive the luminosity of the cluster and of
the central galaxy  (bulge and halo),  near  IR observations would  be
particularly useful as the probe of their giant old stellar population
responsible for the  bulk of the enrichment.  At  the same time deeper
X-ray {\it   XMM-Newton} (for external  and  intermediate regions) and
{\it Chandra} (for the innermost regions) observations are required to
map in greater detail metal abundances in clusters.

\begin{acknowledgements}

SDG would like to acknowledge  useful discussions with S.  Borgani, A.
Edge, A.  Ferrara, F.    Gastaldello, L.  Mayer,   T.  Ponman and   A.
Renzini.   This research has  made  use of the NASA/IPAC Extragalactic
Database  (NED)  which is operated by   the Jet Propulsion Laboratory,
California  Institute of Technology,  under contract with the National
Aeronautics and Space Administration.

\end{acknowledgements}

\end{document}